\documentstyle[preprint,aps]{revtex}
\begin{document}
\draft
\preprint{CLNS 95/1374}
\title{Heavy Baryons and Multiquark Exotics in the Large $N_c$ Limit
\footnote{to appear in the proceedings of the Baryons '95 Conference at
Santa Fe, New Mexico.}}
\author{Chi-Keung Chow}
\address{Newman Laboratory of Nuclear Studies, Cornell University, Ithaca,
NY 14853.}
\date{\today}
\maketitle
\begin{abstract}
In the large $N_c$ limit, one can describe normal heavy baryons $Qqq$ and
heavy pentaquarks $\bar Qqqqq$ as bound states of heavy mesons to chiral
(anti)solitons.
In this picture, the strong and electromagnetic decay parameters of these
hadrons can be calculated from those of the constituent heavy mesons and
light baryons, while the weak decay form factors can be evaluated
analytically.
\end{abstract}
\narrowtext
In the paper, the applications of the bound state picture, a phemenological
model of heavy baryons and multiquarks motivated by the consideration of the
large $N_c$ limit, is briefly reviewed.
In this picture, heavy baryons merge as stable bound state of heavy mesons
to chiral solitons, the binding potential given by chiral perturbation
theory.
Moreover, multiquark exotics can be accomodated in a natural generalization
of our formalism.

The chiral soliton model in the large $N_c$ limit \cite{1} studied the
possibility of identifying baryons as topological solitons in a non-linear
sigma model.
For example, consider two-flavor QCD where chiral SU(2)$_L\times$SU(2)$_R$ is
spontaneously broken into SU(2)$_V$.
The Goldstone bosons, which are the pions, live on the SU(2) manifold.
\begin{equation}
U=\exp\left({2i\tau^a\pi^a\over f}\right)\in {\rm SU(2)},
\end{equation}
where the $\tau^a$ are the SU(2) generators and $f$, the pion decay constant.
The pion field $U$ is a mapping from three-dimensional space ${\bf R}^3$ to
the pion field manifold SU(2), which is homeomorphic to ${\bf S}^3$.
On the other hand, by one point compactification at the spatial infinity,
${\bf R}^3$ can also be compactified to ${\bf S}^3$.
As a result, we have
\begin{equation}
U: {\bf S}^3\rightarrow{\bf S}^3.
\end{equation}
The homotopy theory result $\pi_3({\bf S}^3)={\bf Z}$ implies that it is
possible to have topological solitons in this system.
Such topological solitons, when quantized as fermions, will have $I=J=\,$half
integer.
The particular cases $I=J={1\over2}$ and ${3\over2}$ can be identified as the
nucleon N and Delta $\Delta$ respectively.

The exact form of the pion configuration is given by the hedgehog ansatz,
\begin{equation}
U({\bf x}) = \exp\left({iF(r)\tau_ax_a\over r}\right)
\end{equation}
with the profile function (also called the chiral angle in some literature)
$F(r)$ satisfying the boundary conditions:
\begin{equation}
F(r\rightarrow\infty)=0, \quad F(r=0)=-B\pi,
\end{equation}
where $B$ is the baryon number of the topological soliton.
When $B=1$, the soliton is a baryon and is often called a chiral aoliton in
the literature.

The interaction of a heavy meson with the Goldstone fields by determined by
the chiral lagrangian \cite{2,3,4}.
\begin{equation}
{\cal L}=-i\,{\rm Tr}\,\overline Hv^\mu\partial_\mu H + \textstyle{i\over2}\,
{\rm Tr}\,\overline HHv^\mu U^\dagger\partial_\mu U + \textstyle{i\over2}\,g\,
{\rm Tr}\,\overline HH\gamma^\mu\gamma_5 U^\dagger\partial_\mu U + \dots ,
\label{lag}
\end{equation}
where the ellipsis denotes the contribution of terms containing more
derivatives.
The coupling constant $g$ can be determined by the $D^*\to D\pi$ decay, which
gives $g^2<0.4$ \cite{5}.
Constituent quark model suggests that $g$ is positive, which is crucial to
our discussion.

Since a chiral soliton is nothing but a distribution of the Goldstone fields,
we can investigate the interaction of a heavy meson with a chiral soliton
under the chiral lagrangian above \cite{6,7,8}.
It turns out that the binding potential $V$ is a function of $K=I+s_\ell$,
where $I$ is the isospin and $s_\ell$ is the spin of the light degrees of
freedom of the bound state.
The dependences of $V$ on $I$ and $s_\ell$ enter solely through its dependence
on $K$.
Moreover, when the binding potential is expanded as a Taylor series in
$x$, the relative distance between the heavy meson and (the center of) the
chiral soliton,
\begin{equation}
V(x;K)=V_0(K)+\textstyle{1\over2}\kappa(K)x^2+\dots.
\label{pot}
\end{equation}
it is found that the terms of quartic or higher powers in $x$ of the potential
(the ellipsis in Eq. (\ref{pot})) are subleading in $1/N_c$.
Hence, the potential is {\it exactly simple harmonic} in the large $N_c$
limit.

\begin{mathletters}
When the higher-derivative terms are neglected, the truncated chiral
Lagrangian (\ref{lag}) gives
\begin{equation}
V_0(K=0)=-\textstyle {3\over2} gF'(0),
\end{equation}
and
\begin{equation}
\kappa(K=0)=\kappa=g\left[\textstyle{1\over3}[F'(0)]^3
-\textstyle{5\over6}F'''(0)\right],
\end{equation}
where $F'(0)$ and $F'''(0)$ are respectively the first and third derivative
of the profile function $F(r)$ at $r=0$.
\end{mathletters}
We expect $F'(0)>0$ and $F'''(0)<0$, giving $V_0(K=0)<0$, $\kappa>0$ and
hence stable bound states.
The value of the spring constant $\kappa$ can be determined to be $(530 {\rm
MeV})^3$ in the Skyrme model and $(440 {\rm MeV})^3$ from
$\Lambda^{**}_c-\Lambda_c$ splitting.
The ground states have orbital momentum $L=0$ and $I=s_\ell$.
They can be identified as $\Lambda_Q$, with $I=s_\ell=0$ and $\Sigma^{(*)}_Q$,
with $I=s_\ell=1$.

Under the picture, the strong and electromagnetic properties of a heavy baryon
are given by the properties of its constituents, i.e., the heavy meson and the
chiral soliton.
For example, the $\Sigma_Q\Lambda_Q\pi$ axial current coupling $g_3$ \cite{9}
is calculated in Ref. \cite{6}.
\begin{equation}
g_3=\sqrt{\textstyle{3\over2}}g_A-\sqrt{\textstyle{1\over6}}g,
\end{equation}
where $g_A=1.25$ is the nucleon axial current coupling and $g$ is the
coupling constant appearing in Eq. (\ref{lag}).
As a result, one obatins the estimate
\begin{equation}
\Gamma(\Sigma^{(*)}_Q\to\Lambda_Q\pi)\sim 3.7 \hbox{ MeV}.
\end{equation}
It is also possible to investigate other decay modes like
$\Sigma_c\to\Lambda_c\gamma$ \cite{10} in this picture.

Weak decays $\Lambda_b\to\Lambda_c$ is controlled by the Isgur--Wise form
factor $\eta(w)$.
On the other hand, the $\Sigma^{(*)}_b\to\Sigma^{(*)}_c$ decay is controlled
by two form factors $\zeta_1(w)$ and $\zeta_2(w)$ \cite{11,12,13,14}.
In the bound state picture, it can be proven that they are related \cite{15}
by
\begin{equation}
\zeta_1(w) = -(1+w)\zeta_2 = \eta(w) .
\end{equation}
Moreover, in the large $N_c$ limit, when the binding potential is simple
harmonic, these Isgur--Wise form factors can be analytically evaluated
\cite{8,15}.
\begin{equation}
\eta(w) = \exp\left(-{(w-1)\over2}\sqrt{M_B^3\over\kappa}\right) .
\end{equation}

Similar analysis can also be made on orbitally excited baryons.
Orbitally excited $\Lambda_c$ (denoted by $\Lambda^{**}_c$ in this article)
has been seen in ARGUS \cite{16}, CLEO \cite{17} and E687 \cite{18}.
\begin{mathletters}
In the bound state picture, orbitally excited baryons correspond to excited
states in the binding potential.
The $\Lambda^{**}_c\to\Lambda_c\gamma$ electric dipole transition rate can
be estimated in the bound state picture \cite{19}.
\begin{equation}
\Gamma(\Lambda_c(2593)\to\Lambda_c\gamma)=0.016 \hbox{ MeV},
\end{equation}
\begin{equation}
\Gamma(\Lambda_c(2625)\to\Lambda_c\gamma)=0.021 \hbox{ MeV}.
\end{equation}
\end{mathletters}
The yet unobserved $\Lambda^{**}_b$ are predicted to have masses 5900 and
5926 MeV.
The corresponding E1 decay rate are
\begin{mathletters}
\begin{equation}
\Gamma(\Lambda_b(5900)\to\Lambda_b\gamma)=0.090 \hbox{ MeV},
\end{equation}
\begin{equation}
\Gamma(\Lambda_b(5926)\to\Lambda_b\gamma)=0.119 \hbox{ MeV},
\end{equation}
which may even be the dominant decay mode.
\end{mathletters}
The $\Lambda_b\to\Lambda^{**}_c$ Isgur--Wise form factors (the $\sigma(w)$ in
Ref.~\cite{19.5}) can also be analytically evaluated in this formalism
\cite{20}.

This formalism can also be generalized to incorporate pentaquark exotics by
considering heavy meson--chiral {\it anti}-soliton bound states \cite{21}.
Under the truncated lagrangian (\ref{lag}), the stable states are those with
$K=I+s_\ell=1$, which have binding energy $\tilde V$,
\begin{equation}
\tilde V(K=1)=\textstyle{1\over3}V(K=0).
\end{equation}
Moreover, upon generalization under flavor SU(3) \cite{22}, it can be shown
that the most stable states are exactly the $\bar Qsuud$ and $\bar Qsudd$
states Lipken predicted \cite{23,24}, plus an previously undiscussed
$\bar Qssud$ state.
These states will have $J^P={1\over2}^+$ and their masses can be estimated
in the bound state picture.
\begin{mathletters}
\begin{equation}
|\bar csuud\rangle = |\bar csudd\rangle \sim 2857 \hbox{ MeV},
\end{equation}
\begin{equation}
|\bar cssud\rangle \sim 3009 \hbox{ MeV}.
\end{equation}
\end{mathletters}
These states, if exist, may be seen in the Fermilab experiment E791 in the
near future.
Moreover, the weak decay form factor $\bar bqqqq \to \bar cqqqq$ can also be
analytically evaluated in this framework \cite{21}.

Further extention of this framework in describing tetraquark and hexaquark
exotics are discussed in Ref.~\cite{21,25}.

We conclude by noting that the bound state picture provide us the
possibility of calculating non-perturbative quantities in hadron physics.
The results presented above are in the leading order of $1/M_Q$ and $1/N_c$.
Since both of these parameters are not extremely small, there may be
important $1/M_Q$ and $1/N_c$ corrections.
Another potentially dangerous source of correction are the terms with more
than one derivative in the chiral lagrangian, which are simply ignored in
our discussion.
Further investigations of these higher order corrections will be useful to
the studies of hadron physics.

\end{document}